\documentclass[12pt]{article}
\usepackage{amsmath}
\usepackage{amstext,amssymb,amsfonts}
\usepackage[dvips]{graphicx}
\usepackage{color}
\advance\voffset by -1.5cm
\advance\hoffset by -1.25cm
\def\Im{\,{\rm Im}\, }

\def\be{\begin{equation}}
\def\ee{\end{equation}}
\def\ba{\begin{eqnarray}}
\def\ea{\end{eqnarray}}

\newcommand{\ie}{{\it i.e.~}}

\begin{document}

\vspace*{-1.5cm}
\thispagestyle{empty}
\begin{flushright}
AEI-2007-098
\end{flushright}
\vspace*{2.5cm}

\begin{center}
{\Large 
{\bf Symmetries of perturbed conformal field theories}}
\vspace{2.5cm}

{\large Stefan Fredenhagen$^{1}$}
\footnotetext{$^{1}${\tt E-mail: stefan@aei.mpg.de}},
{\large Matthias R.\ Gaberdiel$^{2}$}
\footnotetext{$^{2}${\tt E-mail: gaberdiel@itp.phys.ethz.ch}} 
{\large and} 
{\large Christoph A.\ Keller$^{3}$}
\footnotetext{$^{3}${\tt E-mail: kellerc@itp.phys.ethz.ch}} 
\vspace*{0.5cm}

$^{1}$Max-Planck-Institut f{\"u}r Gravitationsphysik,
Albert-Einstein-Institut\\ 
D-14424 Golm, Germany\\
\vspace*{0.5cm}

$^{2,3}$Institut f{\"u}r Theoretische Physik, ETH Z{\"u}rich\\
CH-8093 Z{\"u}rich, Switzerland\\
\vspace*{3cm}

{\bf Abstract}
\end{center}
The symmetries of perturbed conformal field theories are 
analysed. We explain which generators of the chiral algebras of a bulk
theory survive a perturbation by an exactly marginal bulk field. We  
also study the behaviour of D-branes under current-current
bulk deformations. We find that the branes always continue to preserve
as much symmetry as they possibly can, {\it i.e.}\ as much
as is preserved in
the bulk. We illustrate these findings with several examples,
including permutation branes in WZW models and B-type D-branes in
Gepner models.


\newpage
\renewcommand{\theequation}{\arabic{section}.\arabic{equation}}


\section{Introduction}
\setcounter{equation}{0}

Chiral symmetry algebras play an important role in the construction of
exactly solvable conformal field theories (see {\it e.g.}\
\cite{Moore:1988qv,Bouwknegt:1992wg,Fuchs:2002cm}). Often these
symmetries are only present at specific points in the closed string
moduli space and are partially broken when the theory is deformed away
from these special points. An important example are Gepner models
which are rational conformal field theories with $N=2$ supersymmetry
at a special point in the moduli space of Calabi-Yau compactifications
of string theory. Starting from such a highly symmetric point, one can
explore the moduli space by perturbing the original theory by some
marginal operator~(see {\it e.g.}\
\cite{Narain:1985jj,Narain:1986am,Ginsparg:1987eb,Seiberg:1988pf,
Kutasov:1988xb}). In this context it is important to know which part
of the symmetry algebra survives the deformation. 

In many situations of interest the string background also involves
D-branes. It is then equally important to understand how much symmetry 
the branes continue to preserve under the bulk deformation. In
this paper we shall give a fairly comprehensive analysis of this
question. Roughly speaking we shall find that branes always continue 
to preserve as much of the original symmetry as is possible, 
{\it i.e.}\ as is preserved in the bulk. We shall also give 
a criterion for which generators of the chiral algebras of the bulk  
survive the perturbation. 
\smallskip

The operators that describe changes of the closed string moduli are
exactly marginal bulk operators. In the presence of boundaries these
operators may cease to be exactly marginal
\cite{Fredenhagen:2006dn}. If this is the case, the bulk perturbation
breaks the conformal invariance of the boundary condition and induces
a renormalisation group flow on the boundary. In general the resulting
boundary condition is then `far away' from the original boundary
condition and it will be difficult to analyse the symmetries it
continues to preserve. We shall therefore always assume that no such
renormalisation group flow will be induced, {\it i.e.}\ that the
boundary theory can be smoothly adjusted to the deformation of the
bulk. This is equivalent to the statement that the boundary condition
continues to be conformally invariant, {\it i.e.}\ that it continues
to satisfy $T(z)=\bar T(\bar z)$ on the boundary of the upper half
plane.

But even if the conformal symmetry is maintained, other symmetries may be
broken by the deformation. Let us assume that in the unperturbed
theory the boundary preserves a chiral symmetry algebra
$\mathcal{A}$. This means that the holomorphic fields
$S\in\mathcal{A}$ are related to the anti-holomorphic fields
$\bar{S}\in\bar{\mathcal{A}}$ on the boundary of the upper half plane
by 
\begin{equation} \label{basic}
\omega(S)(z)=\bar S(\bar z) \ .
\end{equation} 
Here $\omega$ is an automorphism of the symmetry algebra $\mathcal{A}$
that describes how the left- and right-moving chiral algebras are
glued together at the boundary; we have also assumed that the 
anti-chiral symmetry algebra $\bar{\mathcal{A}}$ is isomorphic to 
$\mathcal{A}$. If $\mathcal{A}$ continues to be a
chiral symmetry upon perturbation we can ask whether the gluing
condition (\ref{basic}) is violated or deformed by the
perturbation.\footnote{For boundary perturbations a similar problem
was studied
in~\cite{Callan:1994ub,Polchinski:1994my,Recknagel:1998ih}.} For the
case of a current-current deformation of the bulk we shall see that 
(\ref{basic}), with a suitably modified $\omega$, will always continue
to hold even after the perturbation. Thus boundary conditions always
continue to preserve as much symmetry as they possibly can.

\begin{figure}
\begin{center}
\input{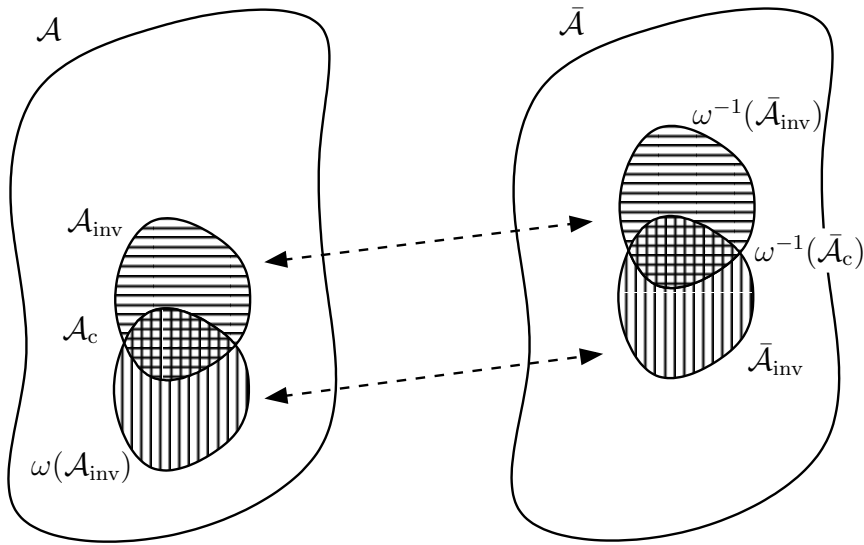}
\end{center}
\caption{\label{fig:algebras}An illustration of the gluing of the
left- and right-moving chiral algebras: By the gluing
condition~\eqref{basic}, the subalgebra
$\bar{\mathcal{A}}_{\text{inv}}$ is glued to
$\omega(\mathcal{A}_{\text{inv}})$, whereas $\mathcal{A}_{\text{inv}}$
is glued to $\omega^{-1} (\bar{\mathcal{A}}_{\text{inv}})$.
After the deformation, only the fields in $\mathcal{A}_{\text{inv}}$
and $\bar{\mathcal{A}}_{\text{inv}}$ stay chiral, which means that it
only makes sense to glue the fields in $\mathcal{A}_{\text{c}}$ and
$\omega^{-1} (\bar{\mathcal{A}}_{\text{c}})$.}
\end{figure}
There is however an interesting subtlety that deserves a comment. If
the bulk perturbation breaks $\mathcal{A}$ down to
$\mathcal{A}_{\text{inv}}$, it only makes sense (see
figure~\ref{fig:algebras}) to require that (\ref{basic}) continues to
hold for fields in\footnote{Here we assume that the left- and
  right-moving algebras are broken in the same way, 
as is for example the case for a current-current deformation by 
$\Phi =J\bar{J}$.} 
\begin{equation}
\mathcal{A}_{\text{c}}=\mathcal{A}_{\text{inv}} \cap
\omega(\medskip\mathcal{A}_{\text{inv}})\ .
\end{equation}
The above argument then only implies that the boundary theory
preserves $\mathcal{A}_{\text{c}} \subseteq
\mathcal{A}_{\text{inv}}$. However, as we shall see, the actual
symmetry of the boundary theory is in fact $\mathcal{A}_{\text{inv}}$:
the spectrum of boundary fields always contains an algebra of mutually
local boundary fields associated to $\mathcal{A}_{\text{inv}}$, and
thus the full spectrum can be organised in (twisted) representations
of $\mathcal{A}_{\text{inv}}$. The subalgebra $\mathcal{A}_{\text{c}}
\subseteq \mathcal{A}_{\text{inv}}$ also plays a special role: it
consists precisely of those fields that are actually local with
respect to all other boundary fields; with respect to
$\mathcal{A}_{\text{c}}$ the full boundary spectrum then forms a
conventional (untwisted) representation.

We shall exemplify these findings (and subtleties) with a number
of examples, most notably branes on a torus and in a product of
$SU(2)$ WZW models. For these cases our results agree with the
geometric intuition about the behaviour of branes under bulk
deformations. We also consider complex structure deformations of
B-type branes in Gepner models for which the perturbing field is not
of current-current type. This problem can be conveniently studied
using the language of matrix factorisations; the results we obtain are
in nice agreement with the expectations based on our general
analysis of current-current deformations.  
\medskip

The paper is organised as follows. In the remainder of this section we
introduce some basic notation. In section~2 we study the preserved
symmetries of a theory without boundaries. We show that a necessary
criterion that 
a chiral field $S$ stays chiral is that $S$ does not have a
simple pole with the perturbing field $\Phi$, see (\ref{chiralcond}).
If $\Phi$ is an exactly marginal current-current field, this is also
sufficient to all orders. 

In section~3 we then consider conformal field theories with
boundary. Any boundary condition must preserve at least the
conformal symmetry, \ie $T(z)=\bar T(\bar z)$ on the boundary. By
investigating the perturbed version of this condition we rederive one
of the key results of \cite{Fredenhagen:2006dn} by a different
method. We then study the behaviour of gluing automorphisms $\omega$ 
under bulk perturbations. Under some weak assumptions we find that it
is always possible to adjust the gluing map, as long as the bulk 
symmetries in question are preserved.

In section~4 we analyse the behaviour of the boundary symmetry algebra
under current-current bulk perturbations directly from 
the point of view of the boundary theory. We derive
all-order criteria that describe when a boundary field 
(i) does not change its conformal weight, when 
(ii) it remains self-local, when 
(iii) two boundary fields stay mutually local, and when 
(iv) a boundary field remains relatively local with respect to all the 
other boundary fields. In particular this gives a criterion for the
existence of a symmetry algebra on the boundary in the sense of an
algebra of mutually local fields. As we shall explain, the spectrum of
boundary fields can then be decomposed into twisted representations of
this algebra. We also discuss a number of examples and comment on the
implications of these results for the structure of the open string
moduli space.

Finally, section~5 contains our conclusions. There are two appendices
where some of the more technical calculations are described.

\subsection{Setup and notation}

Let us begin by reviewing some basic facts about perturbed conformal
field theories; they were first considered
by~\cite{Zamolodchikov:1986gt,Ludwig:1987rk,Ludwig:1987gs,Cardy:1988tj},
for an introduction see for example \cite{Cardy}.  We consider a
conformal field theory with action $S_0$ and perturb the theory by a
marginal operator $\Phi$, where $h_\Phi=\bar h_\Phi$=1. This
perturbation changes the action by
\begin{equation}
\Delta \mathcal{S} = -\lambda \int d^{2}w \, \Phi (w,\bar{w}) \ .
\end{equation}
Here $\lambda$ is a dimensionless coupling constant, and the measure
of the integral is normalised such that $d^{2}w = dx\,dy$ where
$w=x+iy$. Let $\langle \ldots \rangle$ denote the correlators in the
unperturbed theory; the perturbed correlators are then defined as 
\begin{equation}
\langle \phi_{1} (z_{1},\bar{z}_{1})\dotsb \phi_{n}
(z_{n},\bar{z}_{n})\rangle_{\lambda} = \frac{\langle \phi_{1}
(z_{1},\bar{z}_{1})\dotsb \phi_{n} (z_{n},\bar{z}_{n}) e^{-\Delta
\mathcal{S}}\rangle}{\langle e^{-\Delta \mathcal{S}}\rangle} \ ,
\end{equation}
and similarly for correlators involving boundary fields. The
expression on the right hand side is divergent and has to be
regularised. To do so we 
expand the exponential as a power series 
\begin{equation}
e^{-\Delta \mathcal{S}} = \sum_{n=0}^{\infty} \frac{\lambda^{n}}{n!} \int
d^{2}w_{1} \dotsb \int d^{2}w_{n} \, \Phi (w_{1},\bar{w}_{1}) \dotsb \Phi
(w_{n},\bar{w}_{n}) \ ,
\end{equation}
and use the prescription that the domain of integration is
restricted to
\begin{equation}
|w_{i}-w_{j}|>\epsilon \qquad |w_{i}-z_{j}|>\epsilon \ . 
\end{equation}
If we consider a theory on the upper half plane $\mathbb{H^+}$, the
integrals are further constrained by $\Im
w_{i}>\frac{\epsilon}{2}$. The parameter $\epsilon$ is a UV cut-off,
and from the dependence of the correlators on $\epsilon$ one can
derive the renormalisation group equations for $\lambda$ and 
any other couplings (see~\cite{Fredenhagen:2006dn} for an
analysis of the coupled bulk-boundary RG equations). 

We are only interested in deformations $\Phi$ which remain marginal
also in the perturbed theory at finite $\lambda$. Such deformations
are called exactly marginal and appeared first in the analysis of
critical lines of models at central charge~$1$, see {\it e.g.}\
\cite{Kadanoff:1978pv,Cardy:1987vr,Saleur:1987}. They are
characterised by the property that the corresponding coupling constant
$\lambda$ does not run under the renormalisation group flow. A simple
class of exactly marginal operators (which shall be the main focus of
our analysis) are current-current deformations $\Phi= J \bar{J}$
with currents $J$ and $\bar{J}$. In the presence of a boundary we
shall also assume that $J=\omega(\bar{J})$ at the boundary. The
perturbation $\Phi$ is then exactly marginal on worldsheets with
boundary if the OPE of $J$ and $\omega(J)$ does not contain a marginal
or relevant field~\cite{Fredenhagen:2006dn}. Some of our statements
also generalise to more general current-current deformations of the
form $\Phi=\sum J_{i}\bar J'_{i}$; this will be briefly
discussed in the conclusions (see section~4).

Throughout the paper we shall assume that the spectrum of $L_0$ and
$\bar{L}_0$ is bounded from below by zero, and that the state with
$L_0=\bar{L}_0=0$ is unique (the vacuum). Then it follows for example
that the exactly marginal bulk field $\Phi$ is a primary field with
respect to both Virasoro algebras.


\section{Bulk symmetries}
\setcounter{equation}{0}

First we want to investigate under which conditions bulk symmetries
are preserved by an exactly marginal bulk deformation.  Let us assume
that the field $S$ belongs to the chiral symmetry algebra
$\mathcal{A}$ before the perturbation. We want to ask whether the
corresponding symmetry is preserved under the deformation, \ie whether
$S$ remains a chiral field. Equivalently we can investigate
whether the operator product expansion with the anti-holomorphic
stress-energy tensor $\bar{T}$ remains trivial.

In order to study this question we consider the correlator 
$\langle \dotsb S (z) \bar{T} (\bar{v}) \rangle$ 
and look for singularities in $\bar{z}-\bar{v}$. To first order 
in the perturbation we find 
\begin{align}\label{eq1}
\lambda  \int d^{2}w & \, \langle \dotsb S (z) \bar{T} (\bar{v}) \Phi
(w,\bar{w})\rangle \nonumber\\
 & \sim \ \lambda \int d^{2}w \, \Big(\frac{1}{(\bar{v}-\bar{w})^{2}} 
+ \frac{1}{\bar{v}-\bar{w}}\partial_{\bar{w}} \Big) 
 \langle \dotsb S (z) \Phi (w,\bar{w})\rangle\nonumber\\
&\sim \ \lambda \int d^{2}w \, \partial_{\bar{w}}
\Big(\frac{1}{\bar{v}-\bar{w}}
\langle \dotsb S (z)\Phi (w,\bar{w})\rangle \Big) \nonumber\\ 
&\sim \ \frac{\lambda}{2i} \oint_{|z-w|=\epsilon} dw \
\frac{1}{\bar{v}-\bar{w}} \langle \dotsb S (z)\Phi (w,\bar{w})\rangle
\ .
\end{align}
Here we have only kept the terms that could contribute to a
singularity in $\bar{v}-\bar{z}$. From the last expression we see that
--- in the limit $\epsilon \to 0$ --- we get a contribution precisely if
there is a term proportional to $(w-z)^{-1}$ (expanded in $w$ around
$z$); in that case we find a singularity proportional to
$(\bar{v}-\bar{z})^{-1}$. [One might have expected that there could
also be a term $(\bar{v}-\bar{z})^{-2}$ which in the operator product
expansion (OPE) would correspond to a non-vanishing right-moving
conformal weight $\bar{h}$, but such a term can only arise at second
order.]

We can therefore conclude that the chiral symmetry $S$ is preserved to
first order in the perturbation if
\begin{equation}\label{chiralcond}
\lim_{\epsilon \to 0} \oint_{|w-z|=\epsilon } dw \ 
\Phi (w,\bar{w}) S(z) = 0 \ .
\end{equation}
By assumption this condition must be satisfied if we take $S$ to be
the chiral component of the stress energy tensor $S=T$. Indeed, since 
$\Phi$ is marginal we have 
\begin{equation}
\Phi (w,\bar{w}) T(z) = \frac{1}{(z-w)^2} \, \Phi(w,\bar{w}) + 
\frac{1}{z-w} \partial_w \Phi(w,\bar{w}) + {\cal O}(1)  \ , 
\end{equation}
from which (\ref{chiralcond}) immediately
follows.\footnote{In~\cite{Sen:1990hh} the preservation of conformal
symmetry was demonstrated by showing that the modes of the deformed
energy-momentum tensor still form a Virasoro algebra with the same
central charge.}

The condition (\ref{chiralcond}) applies to general marginal
deformations $\Phi$. In the specific case where $\Phi$ is a 
current-current deformation, $\Phi(w,\bar{w})=J(w)\bar{J} (\bar{w})$, the
above condition simply amounts to the requirement that $S$ does not
carry any charge corresponding to $J$, so the symmetry algebra after
deformation is 
$\mathcal{A}_{\text{inv}}=\{S\in \mathcal{A}, J_{0}S =0\}$. Similarly,
the unbroken anti-chiral symmetry algebra after the deformation is 
$\bar{\mathcal{A}}_{\text{inv}}=\{\bar{S}\in \bar{\mathcal{A}}, 
\bar{J}_{0}\bar{S} =0\}$.

\subsection{Higher order analysis}

The above analysis has only been to first order in the perturbation,
and the condition (\ref{chiralcond}) is therefore only a necessary
condition. In order to analyse whether it is also sufficient we have
to check whether \eqref{chiralcond} remains true even after
perturbation. If this is the case, then at every order in
perturbation theory we can use the above argument to deduce that we
do not obtain a non-trivial OPE between $\bar{T}$ and $S$.  

Unfortunately, the analysis is quite complicated in the general case,
but we can say something for the special case where the perturbation
is a current-current deformation, $\Phi =J\bar{J}$. We normalise the
currents so that 
\begin{equation}
J(z) J(w) = \frac{1}{(z-w)^2} + {\cal O}(1) \ , \qquad 
\bar{J}(\bar{z}) \bar{J}(\bar{w}) = \frac{1}{(\bar{z}-\bar{w})^2} 
+ {\cal O}(1) \ . 
\end{equation}
The condition~\eqref{chiralcond} implies that the OPE of $J$ and $S$
is of the form  
\begin{equation}\label{assu}
J (w) S (z) = \sum_{n=1}^{h_{S}}\frac{1}{(w-z)^{n+1}} 
V (J_{n}S,z) + {\cal O}(1) \ ,
\end{equation}
{\it i.e.}\ that the $n=0$ term is not present. Here $V(\phi,z)$
denotes the field corresponding to the state $\phi$. We want to check
whether the property~\eqref{chiralcond} remains true to next order in
perturbation theory. To this end we consider the correlator
\begin{equation}
I=\langle \dotsb  \Phi (w,\bar{w}) \, S (z) \int d^{2}v\, \Phi (v,\bar{v})
\rangle \ ,
\end{equation}
and look for a term corresponding to a simple pole $(w-z)^{-1}$. The
integrand can be calculated by expressing $\Phi$ in terms of the
currents $\Phi=J\bar{J}$, and by 
using the operator product expansion of
$\bar{J}(\bar{w})$ with the other fields; this leads to 
\begin{align}\label{integral}
I& = \int d^{2}v \, \left\langle \dotsb J (w)S (z) J (v)
\frac{1}{(\bar{v}-\bar{w})^{2}}\right\rangle \\
\label{integral1} & \quad  +\int d^{2}v\, \left\langle \Big( 
\sum_{i} \dotsb
\left[ \bar{J}(\bar{w}) \phi_i (z_i,\bar{z}_i) \right]   
\dotsb \Big) J(w) S(z) J(v) \bar{J}(\bar{v})\right\rangle \ ,
\end{align}
where we have denoted the singular contribution of the OPE of
$\bar{J}(\bar{w})$ with the field $\phi_i (z_i,\bar{z}_i)$ by a square
bracket. By the same trick as in~\eqref{eq1}, the first integral can
be written as a contour integral over $dv$ which encircles the points
$w,z$, as well as the other insertion points $z_{i}$,
\begin{equation}
\oint_{w,z,z_{i}} dv \, \langle \dotsb J (w) S (z) J (v) \rangle
\frac{1}{\bar{v}-\bar{w}} \ .
\end{equation}
The contribution around $w$ does not have any singularity in $w-z$,
the contribution from $z$ vanishes by the above assumption
(\ref{assu}), and the contribution from the $z_{i}$ cannot generate
any new pole in $(w-z)$. 

In the second integral \eqref{integral1}, we replace $J(w)$ again by
the singular terms of its operator product expansions. The
contributions from the insertion points $z_{i}$ and $v$ do not have
any singularity in $w-z$, and the contribution from $z$ cannot produce
a single pole $(w-z)^{-1}$ again because of (\ref{assu}).

Thus we can conclude that the condition~\eqref{chiralcond} also holds 
in the deformed theory, and the chirality of $S$ is not spoiled at
second order. Assuming that $J$ and $\bar{J}$ remain currents along
the deformation, this shows that the chirality is preserved
to all orders in perturbation theory.

\subsection{An example: Gepner models}

To illustrate the general condition (\ref{chiralcond}) let us consider
a Gepner model~\cite{Gepner:1987qi} corresponding to the Calabi-Yau
$3$-fold $W=0$ in weighted complex projective space, where
\begin{equation} 
W= \sum_{i=1}^{5} x_i^{n_i} \ , \qquad \hbox{and} \qquad
\sum_{i=1}^{5} \frac{1}{n_i} = 1 
\ . 
\end{equation}
The relevant Gepner model is then (an orbifold of) the tensor product
of five $N=2$ superconformal minimal models at level $k_i=n_i-2$. The
corresponding bulk conformal field theory possesses the diagonal $N=2$
superconformal symmetry at central charge $c=9$, but also the five
individual $N=2$ superconformal symmetries at
\begin{equation}
c_i = \frac{3 k_i}{k_i+2} \ , \qquad n_i=k_i+2 \ . 
\end{equation}
Complex structure deformations of the Calabi-Yau manifold are
described by polynomials in $x_i$ that have the appropriate scaling
behaviour in the weighted projective space (see {\it e.g.}\ the
review~\cite{Greene:1996cy}); in terms of conformal field theory,
these deformations are described by the (cc) fields
\begin{equation}\label{cc}
\hat\Phi = \phi^1_{l_1l_10}\ \phi^2_{l_2l_20}\
\phi^3_{l_3l_30}\ \phi^4_{l_4l_40} \ \phi^5_{l_5 l_5 0} \ , \qquad
\hbox{where} \qquad
\sum_{i=1}^{5} \frac{l_i}{k_i+2} = 1 \ . 
\end{equation}
Here $\phi^i_{lms}$ denotes the bulk field corresponding to the
representation $(lms)\otimes\overline{(lms)}$ of the $i^{\rm th}$
$N=2$ factor, and $(lms)$ are the usual coset labels of the $N=2$ 
superconformal algebra --- see for example~\cite{Greene:1996cy}.   
The field $\hat\Phi$ has total conformal weight equal to $h=1/2$, and
total $u(1)$ charge $1$, $J_0 \hat\Phi=\hat\Phi$. 

The actual perturbing field is obtained from $\hat\Phi$ by application
of the $N=1$ supercharges $G=G^++G^-$ and
$\bar{G} = \bar{G}^+ + \bar{G}^-$. Since $\hat\Phi$ is a cc field, 
$G^+_{-1/2}\hat\Phi = \bar{G}^+_{-1/2}\hat\Phi=0$, and hence the
perturbing  field is $\Phi\equiv G^-_{-1/2}\bar{G}^-_{-1/2}\hat\Phi$. 
Since $\hat\Phi$ has total $u(1)$ charge equal to $q=\bar{q}=1$, it
follows that $\Phi$ has total $u(1)$ charge equal to zero, 
\begin{equation}\label{charge}
J_0 \, \Phi = 0 \ ,
\end{equation}
and similarly $\bar{J}_0 \, \Phi = 0$. 

We can now apply our condition (\ref{chiralcond}) to the different
generators of the diagonal $N=2$ superconformal algebra. Since
$\Phi$ is (exactly) marginal, the conformal symmetry is preserved
(see the discussion following (\ref{chiralcond})). As for the total
$u(1)$ current of the $N=2$ algebra, taking $S=J$, (\ref{chiralcond})
vanishes because of (\ref{charge}). Finally, the
operator product expansion of $G^+$ with $\Phi$ is a total derivative:
let $\tilde{\Phi}=\bar{G}^{-}_{-1/2}\hat{\Phi}$, then the OPE is
\begin{align}
G^+(z)\Phi(w,\bar w) &\sim \frac{1}{(z-w)^2} V(G^+_{1/2}
G^-_{-1/2} \tilde
\Phi , w,\bar{w}) + \frac{1}{z-w}
V(G^+_{-1/2}G^-_{-1/2}\tilde \Phi, w,\bar{w})  \nonumber\\
& \sim  \frac{1}{(z-w)^2} V((2L_0+J_0)\tilde \Phi,
w,\bar{w})
+ \frac{1}{z-w} V(2L_{-1} \tilde \Phi, w,\bar{w}) \nonumber\\
& \sim 
2\,\partial_w\left( \frac{1}{z-w} V(\tilde
\Phi,w,\bar{w}) \right)\ ,
\end{align}
so that the residue integral (\ref{chiralcond}) is zero. On the other
hand, the OPE with $G^-$ vanishes directly.
It thus follows that the diagonal $N=2$ superconformal algebra remains
a symmetry under this perturbation. This is certainly what one expects
since the perturbation preserves spacetime supersymmetry.

On the other hand, it follows from a similar reasoning that the
$i^{\rm th}$ $N=2$ superconformal symmetry is not preserved unless
$l_i=0$. This is obviously also in line with general expectations.

\subsection{Another example: WZW models and the free boson}

As another example we consider a Wess-Zumino-Witten (WZW) model
describing strings on a Lie group $G$. The symmetry algebra
$\mathcal{A}$ is generated from the affine Lie algebra $g$
corresponding to $G$. A current-current deformation $\Phi =J\bar{J}$
singles out a subgroup $U(1)\subset G$ (for a discussion of
current-current deformations of WZW models
see~\cite{Hassan:1992gi,Forste:2003km}). We can decompose the vacuum
sector in terms of representations of the coset algebra $g/u(1)$ and
the $u(1)$ theory,\footnote{Here the $u(1)$ algebra is just the
Heisenberg algebra of one current, without the inclusion of any
charged fields.}  
\begin{equation}
\mathcal{H}^{g}_{0} = \bigoplus_{m\in\mathbb{Z}} 
\mathcal{H}_{(0,m)}^{g/u(1)} \otimes
\mathcal{H}^{u(1)}_{m} \ .
\end{equation}
Projecting onto the states which are uncharged with respect to the
$u(1)$ leaves us with the term with $m=0$. Thus the
subalgebra $\mathcal{A}_{\text{inv}}$ that is preserved under the
deformation is precisely the tensor product of the coset algebra and
the $u(1)$ algebra.

A particular example is the $SU (2)$ WZW model at level $k=1$. It
describes the same theory as the free boson compactified on a circle
at the selfdual radius. The marginal deformations have been
investigated, and the complete connected moduli space is
known~\cite{Ginsparg:1987eb,Dijkgraaf:1987vp}. When the theory is
infinitesimally deformed away from the selfdual radius, the symmetry
is broken to the $u(1)$ algebra. When we continue to deform the
theory, other points of enhanced symmetry will be reached: in fact at
any rational value of the radius squared (when the radius is measured
in units of the selfdual radius) there will be an enhanced
symmetry. Similar considerations apply to the moduli space of $N=1$
supersymmetric theories of a free boson and a free fermion on a
circle~\cite{Dixon:1988ac}.


\section{Deformed gluing conditions}
\setcounter{equation}{0}

Up to now we have analysed whether symmetries of the bulk theory
remain intact under perturbations by exactly marginal bulk operators. We
have seen that a necessary condition for this is
(\ref{chiralcond}). For the case of current-current deformations we 
have furthermore shown that this condition guarantees that $S$ remains
chiral to arbitrary order in perturbation theory.  

Now we want to analyse how symmetries of a boundary theory are
affected by a bulk perturbation. To this end we introduce a boundary
into our theory and consider the conformal field theory on the upper
half plane $\mathbb{H}^{+}$. In general, an exactly marginal bulk
perturbation can break the conformal invariance at the boundary and
induce a renormalisation group flow \cite{Fredenhagen:2006dn}. If this
is the case, it will be very difficult to make predictions about the
symmetries of the fixed point theory (since the fixed point will be,
in general, far away from the original boundary theory). We shall
therefore restrict ourselves to bulk deformations which do not induce
a non-trivial RG flow, and which therefore preserve the conformal
invariance on the upper half plane. This will be the case provided
that certain bulk-boundary OPE coefficients vanish
\cite{Fredenhagen:2006dn}; we shall give an independent proof of this
result in the following subsection.

If this condition is satisfied, we can ask how the symmetries of a
boundary theory (that come from bulk symmetries) behave under a bulk
perturbation. We shall give arguments to suggest that the boundary
condition always preserves those gluing conditions that continue to
make sense in the bulk.

\subsection{Preserving the conformal invariance}

Let us begin by analysing whether the boundary condition remains
conformally invariant under a bulk perturbation. Conformal invariance
of a boundary condition requires that the energy
momentum tensor satisfies the gluing condition
\begin{equation}\label{conglue}
T (z) = \bar{T} (\bar{z}) \quad \text{at}\ z=\bar{z}\ .
\end{equation}
We want to study whether this condition remains true under the bulk
perturbation, so we have to look at the limit of correlators
\begin{equation}
\lim_{y\to 0}\langle \dotsb (T (z)-\bar{T} (\bar{z}))\rangle_{\lambda}
\end{equation}
with $z=x+iy$. The first order correction to the gluing condition comes from 
\begin{equation}
\Delta T = \lim_{y\to 0} \lambda \int_{\text{Im}\, w>\epsilon/2}d^{2}w\, 
\big(T(z)-\bar{T} (\bar{z}) \big)\Phi (w,\bar{w}) \ ,
\end{equation}
where the expression is understood to be inserted into a
correlator. Note that it is important that we first make a Laurent
expansion in the regulator $\epsilon$ before taking the limit 
$y\to 0$, as otherwise the expression vanishes trivially. Since $\Phi$
is primary, the singular part of the OPE in the presence of the
boundary is 
\begin{align}
& \big(T(z)-\bar{T} (\bar{z}) \big)\Phi (w,\bar{w})  \sim  
\frac{1}{(z-w)^2} \, \Phi (w,\bar{w}) 
+ \frac{1}{(z-w)} \, \partial_w \Phi(w,\bar{w}) -
((z,w)\rightarrow (\bar{z},\bar{w})) \nonumber \\
& \qquad \qquad \qquad \qquad \qquad \quad
+ \frac{1}{(z-\bar{w})^2} \, \Phi (w,\bar{w}) 
+ \frac{1}{(z-\bar{w})} \partial_{\bar{w}} \, \Phi(w,\bar{w}) -
((z,\bar{w})\rightarrow (\bar{z},w)) \ , \nonumber
\end{align}
where the second line arises from the mirror images that are required
to guarantee that (\ref{conglue}) holds in all correlators. As before
in (\ref{eq1}) we can rewrite the right hand side in terms of
derivatives with respect to $\partial_w$ and $\partial_{\bar{w}}$;
we can thus write $\Delta T$ as 
\begin{align}\label{eq2}
\Delta T & =  \lim_{y\to 0} \Biggl[ \frac{i\lambda}{2} \int_{\text{Im}\,
w=\epsilon/2} d\bar{w}\, \bigg\{ 
\frac{1}{z-w} \Phi(w,\bar{w}) - (z\to \bar{z})  \bigg\} \nonumber\\
& \qquad \qquad 
- \frac{i\lambda}{2} \int_{\text{Im}\, w=\epsilon/2} dw \, \bigg\{
\frac{1}{z-\bar{w}} \Phi(w,\bar{w}) - (z\to \bar{z}) \bigg\} \Biggr]
\nonumber\\
& = \lim_{y\to 0} \frac{i\lambda}{2} \int_{\mathbb{R}} du \,\bigg\{
\frac{1}{z-u-i\epsilon/2} - \frac{1}{z-u+i\epsilon/2} - (z\to \bar{z})
\bigg\} \, \Phi(u+i\epsilon/2;u-i\epsilon/2) \nonumber\\
& = \lim_{y\to 0} \frac{i\lambda}{2} \int_{\mathbb{R}} du \,\bigg\{ 
\frac{i\epsilon}{(z-u)^{2} + \epsilon^2/4} - (z\to \bar{z}) \bigg\} \,
\Phi(u+i\epsilon/2 ,u-i\epsilon/2) \ .
\end{align}
Here the minus sign in the second line arises because 
$d^2w = \tfrac{i}{2} \, dw \wedge d\bar{w} 
= - \tfrac{i}{2} \, d\bar{w} \wedge dw$. 
For small $\epsilon$ we can now use the bulk-boundary operator
product expansion to write 
\begin{equation}
\Phi (u+i\epsilon/2 ,u-i\epsilon/2) \sim \sum_{i}
\epsilon^{h_{\psi_{i}}-2} \, B_{i}\, \psi_{i} (u) \ , \label{bubdr}
\end{equation}
where $\psi_{i}$ are boundary fields and the $B_{i}$ are the bulk
boundary coefficients. Since $h_\psi\geq 0$, the most singular term is
proportional to $\epsilon^{-2}$; thus we may drop the $\epsilon^2/4$
term in the denominator of the last line of (\ref{eq2}). The limit 
$y\to 0$ of the bracket in (\ref{eq2}) is of the form
\begin{equation}
\label{deltadistribution}
\lim_{y\to 0} \bigg( \frac{1}{(x+iy-u)^{n}} - \frac{1}{(x-iy-u)^{n}}\bigg) = 
(-1)^{n} \frac{2\pi i}{(n-1)!} \,\delta^{(n-1)} (u-x) 
\end{equation}
with $n=2$, and hence leads to a derivative of a delta
function. Altogether we therefore find  
\begin{equation}
\Delta T = -\pi i \lambda \sum_{i} \epsilon^{h_{\psi_{i}}-1}
B_{i}\partial_{x} \psi_{i} (x) \ . 
\end{equation}
It follows that the gluing condition for $T$ is violated if
in the bulk-boundary operator product expansion of $\Phi$ there are relevant
or marginal boundary fields ($h_{\psi_{i}}\leq 1$). The only exception
is the vacuum ($h=0$), since then there is no $x$-dependence and the
derivative vanishes. This analysis reproduces precisely the 
condition that was found in~\cite{Fredenhagen:2006dn} from considering
the combined bulk-boundary renormalisation group equations. 

In the case of current-current deformations $\Phi=J \bar{J}$, for
which $\bar{J}=\omega(J)$ at the boundary with $\omega(J)$ some chiral
current, the above condition is simply the requirement that the OPE of
$J$ with $\omega(J)$ does not contain a simple pole. In this case the
argument generalises to all orders: it is not difficult to show that
the OPE of $J$ and $\omega(J)$ will not acquire a pole term under the
deformation, so that $\Delta T = 0$ also for finite $\lambda$. For a
general perturbation, however, the first order criterion only provides
a necessary, but not a sufficient condition for $\Phi$ to be exactly
marginal.

\subsection{Preserving a general symmetry}

Let us now assume that the bulk deformation is exactly marginal on
surfaces with boundary so that no relevant or marginal field is
switched on at the boundary. In this case the boundary only adjusts
infinitesimally to the bulk perturbation and we may hope to make
statements about the symmetries it will continue to preserve. 

In the following we shall only consider current-current deformations
$\Phi= J \bar{J}$ for which $\bar{J}=\omega(J)$ at the boundary. Here
$\omega$ is an automorphism of the chiral algebra that is preserved by
the boundary. As we have just explained, in order for this perturbation
to be exactly marginal in the presence of the boundary, we need to have
that the OPE of $J$ with $\omega(J)$ does not contain a simple pole.

Suppose now that the boundary condition preserves the symmetry
associated to some chiral field $S$, 
\begin{equation}\label{Ssym}
\omega (S) (z) = \bar{S} (\bar{z}) \quad \text{at}\ z=\bar{z} \ ,
\end{equation}
where $\omega$ is an automorphism of the preserved chiral algebra
$\mathcal{A}$. We want to ask whether after the perturbation by
$\Phi$, (\ref{Ssym}) still holds, possibly for some adjusted
$\omega$. Obviously for this to make sense we have to require that
$\omega (S)$ continues to be a chiral field even after the
perturbation ($\omega(S)\in \mathcal{A}_{\text{inv}}$), 
and similarly for $\bar{S}$; thus we want to assume that
(\ref{chiralcond}) holds for $\omega (S)$, and similarly 
for $\bar{S}$. Since $\bar{\mathcal{A}}\cong \mathcal{A}$ the latter
condition is equivalent to the statement that $J$ does not have a
simple pole with $S$. Altogether we thus require that the
OPEs of $J$ with $S$ and $\omega(S)$ do not have simple poles. 

There is one subtle point that is worth mentioning. If 
$\omega (\mathcal{A}_{\text{inv}})\not\subset
\mathcal{A}_{\text{inv}}$, the field  
$\omega(S)$ can only be chosen from the intersection
$\mathcal{A}_{\text{c}}=\mathcal{A}_{\text{inv}} \cap \omega
(\mathcal{A}_{\text{inv}})$.  The symmetry algebra on the boundary 
that can arise from gluing bulk fields is then smaller than the
symmetry $\mathcal{A}_{\text{inv}}$ that is preserved in the
bulk. As we shall explain in section~4, the boundary theory actually 
still preserves (in a certain sense) the full symmetry algebra
$\mathcal{A}_{\text{inv}}$; for the time being, however, we
concentrate on the symmetries $\mathcal{A}_{\text{c}}$ that can be
understood in terms of gluing conditions.
\smallskip

Let us thus consider a field $\omega (S)\in\mathcal{A}_{\text{c}}$. As in
the previous subsection we want to study the expression (this is
again to be understood to be inserted into an arbitrary correlator)
\begin{align}
\Delta S & = 
\lim_{y\to 0}  \, \lambda \int d^{2}w\, \big(\omega (S) (z)-\bar{S}
(\bar{z}) \big) \Phi (w,\bar{w}) \nonumber\\
& = \lim_{y\to 0} \lambda \int d^{2}w\, \Big(\sum_{n=1}^{h_{S}}
\frac{1}{(w-z)^{n+1}} V (J_{n}\omega(S),z)\; 
 \bar{J} (\bar{w})\nonumber\\
& \qquad \qquad \qquad\qquad
+\sum_{n=1}^{h_{S}} \frac{1}{(\bar{w}-z)^{n+1}} 
V (\omega(J)_{n}\omega(S),z) \; J (w) \ - (z\to \bar{z}) \Big) 
\label{changeingluingconditions}\\
& = - \lim_{y\to 0} \frac{i\lambda}{2}
\int_{\mathbb{R}} du \, \Big( \sum_{n=1}^{h_{S}}  
\frac{1}{n} \frac{1}{(u-z)^{n}} V (J_{n}\omega(S),z)\; \omega(J)
(u) \nonumber\\
& \qquad \qquad \qquad\qquad - \sum_{n=1}^{h_{S}} \frac{1}{n}
\frac{1}{(u-z)^{n}} V (\omega(J)_{n}\omega(S),z) \; J(u) 
- (z\to \bar{z}) \Big) \ .
\end{align}
We now want to close the contour in the upper half plane. The poles
at insertion points of other fields in the correlator cancel in the
expression in the limit $y\to 0$. The only pole that can give a
contribution is at $u=z$. To determine its residue we use the
full OPE of the fields,  
\begin{equation}
\omega(J) (u) \; V (J_{n}\omega(S),z) = \sum_{m\leq h_{S}-n} 
\frac{1}{(u-z)^{m+1}}V\Bigl(\omega(J)_{m}J_{n}\omega(S),z\Bigr) \ . 
\end{equation}
The residue thus comes from the term with $m=-n$, so that we obtain
\begin{equation}
\Delta S = \pi \lambda  \sum_{n=1}^{h_{S}}\frac{1}{n} \Big[
V \Bigl(\omega(J)_{-n}J_{n}\omega(S),x\Bigr)
- V \Bigl(J_{-n}\omega(J)_{n}\omega(S),x\Bigr) \Big] \ .
\end{equation}
Introducing the operator 
\begin{equation}
K^{\omega} = -i \sum_{n>0} \frac{1}{n} \Big(\omega(J)_{-n}\, J_{n} 
- J_{-n}\, \omega(J)_{n} \Big) \ , 
\end{equation}
we can rewrite the result as 
\begin{equation}
\Delta S = \lim_{y\to 0}\lambda \int d^{2}w\, \big(\omega(S) (z)-\bar{S}
(\bar{z}) \big) \, \Phi (w,\bar{w}) = i\pi \lambda V
(K^{\omega}\omega(S),x) \ .
\end{equation}
This suggests that we can absorb the change $\Delta S$ into a
redefinition of the automorphism $\omega$, {\it i.e.}\ that we have to
first order in the perturbation 
\begin{equation}\label{eq3}
\omega_{\lambda} (S) = \omega (S) -i\pi \lambda V (K^{\omega}\omega (S))\ .
\end{equation}
We need to show that $\omega_{\lambda}$ (to first order in $\lambda$)
still defines an automorphism. Suppose that we can write $J=i \partial
X^1$ and $\omega(J)=i \partial X^2$, where $X^1$ and $X^2$ are free
boson fields. Then $K^\omega$ is precisely the rotation generator in
the $X^1-X^2$ plane, except that in the definition of
$K^\omega$ the zero modes are missing. Since by assumption
$J_0\omega(S)=\omega(J)_{0}\omega(S)=0$, these zero modes can be added
to $K^\omega$ without modifying the action on $\omega(S)$. Thus we can
think of the correction term in (\ref{eq3}) as an infinitesimal
rotation generator, implying that (\ref{eq3}) defines indeed an
automorphism of the chiral algebra.\footnote{As we shall see this is
precisely what happens in the explicit example we are about to
discuss.} More generally, we can prove without any further assumptions
that (\ref{eq3}) defines an automorphism if $S$ is a current; this is 
explained in appendix~A.

\subsection{Example: Diagonal torus branes}

As an example of the results above we consider a diagonal
one-dimensional brane on a square torus with radii $R_{1}=R_{2}=R$,
satisfying the gluing conditions
\begin{equation}\label{diagglue}
J^{1} (z) = \bar{J}^{2} (\bar{z}) \ ,\qquad J^{2} (z)=\bar{J}^{1}
(\bar{z})\ , \qquad z=\bar{z} \ ,
\end{equation}
where $J^{l}=i\partial X^{l}$ is the $u(1)$ current corresponding to
the $l^{\text{th}}$ direction. We now deform the torus by changing
the first of the two equal radii, setting $\Phi =J^{1}\bar{J}^{1}$. 

Obviously this bulk perturbation preserves the chiral
$u(1)$-symmetries, \ie satisfies (\ref{chiralcond}). Furthermore, 
if we take the bulk perturbation $\Phi$ to the boundary we do not
switch on a marginal or relevant field since
\begin{equation}
\left. \Phi(z,\bar{z}) \right|_{y\rightarrow 0} \sim
J^{1}(x+iy) \, J^2(x-iy) \sim {\cal O}(1) \ . 
\end{equation}
Thus the bulk perturbation is exactly marginal on the disk and we
expect that the boundary condition continues to preserve the above
$u(1)$-symmetries. The gluing condition, however, will get
adjusted as detailed above. In fact, one can guess that the adjustment
of the gluing conditions will simply describe the fact that the brane
will continue to stretch diagonally across the torus. This motivates
us to make the ansatz for the gluing conditions
\begin{align}
\omega_{\lambda} (J^{1}) &= -\cos \varphi_{\lambda} \, J^{1} 
+ \sin \varphi_{\lambda}\, J^{2}\\
\omega_{\lambda} (J^{2}) &= \sin \varphi_{\lambda}\, J^{1} 
+\cos \varphi_{\lambda} \, J^{2} \ ,
\end{align}
where $\varphi_{\lambda}$ is a $\lambda$-dependent angle with initial
condition $\varphi_{0}=\frac{\pi}{2}$. Let us fix some value for the
parameter $\lambda$, and consider a small shift $\lambda \to \lambda
+\delta \lambda$. The change in the gluing condition for $J^{1}$ is
then given by
\begin{equation}
i\pi \, \delta \lambda\, V (K^{\omega_{\lambda}}\omega_{\lambda}
(J^{1})) = -\pi \, \delta \lambda (\sin^2 \varphi_{\lambda}\, J^{1} +\sin
\varphi_{\lambda}\, \cos \varphi_{\lambda}\, J^{2}) \ ,
\end{equation}
and similarly for $J^{2}$. If our ansatz is correct, we must be able
to absorb this shift into a redefinition of $\varphi_\lambda$. Thus we
obtain the differential equations for the angle $\varphi_{\lambda}$,
\begin{equation}
\frac{d}{d\lambda} \cos \varphi_{\lambda} = -\pi \sin^{2} \varphi_{\lambda}
\qquad
\frac{d}{d\lambda} \sin \varphi_{\lambda} = \pi \sin  \varphi_{\lambda} \, 
\cos \varphi_{\lambda} \ .
\end{equation}
These are both equivalent to the differential equation for
$\varphi_\lambda$ 
\begin{equation}
\frac{d}{d\lambda}  \varphi_{\lambda} = \pi \sin  \varphi_{\lambda} \qquad
\hbox{or} 
\qquad
\frac{d}{d\lambda}\xi (\lambda) = \pi\xi (\lambda) \ ,
\end{equation}
where $\xi (\lambda) =\tan \frac{\varphi_{\lambda}}{2}$. The solution is 
\begin{equation}
\xi (\lambda) = e^{\pi \lambda} = \tan \frac{\varphi_{\lambda}}{2} \ .
\end{equation}
\begin{figure}
\begin{center}
\input{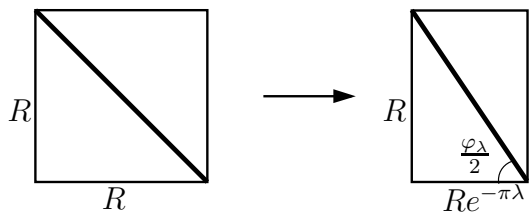}
\end{center}
\caption{\label{fig:torusdef}When one of the radii in the two-torus is
deformed, the brane continues to stretch diagonally and its inclination
changes.}
\end{figure}
The gluing condition $\omega_{\lambda}$ then corresponds to a brane at
angle $\frac{\varphi_{\lambda}}{2}$ which sits on the diagonal of a torus
with radii $R_{1} (\lambda)=R e^{-\pi \lambda}$ and $R_{2}=R$, in
precise agreement with the geometrical intuition (see
figure~\ref{fig:torusdef}).

This is in perfect agreement with the $\lambda$-dependence of the
deformed radius $R_{1} (\lambda)$, as we shall now explain. The 
action for the free boson on the first circle of radius $R_{1}$ has the same
form as the infinitesimal perturbation, 
\begin{equation}
\mathcal{S} +\Delta \mathcal{S} = \frac{1}{2\pi} \int d^{2}w \, \Phi
(w,\bar{w}) -\delta\lambda \int d^{2}w \, \Phi (w,\bar{w})\ ,
\end{equation}
with $\Phi = J_{1}\bar{J}_{1}$ as above. To get back to the standard
normalisation of the action, we have to rescale $J_{1}$ to
$J_{1}'=J_{1} (1-\pi \, \delta\lambda)$ and similarly for
$\bar{J}_{1}$. Since $J=i\partial X_{1}$ and $X_{1}$ has periodicity
$2\pi R_{1}$, the rescaled $X'_{1}$ has periodicity $2\pi R'_{1}$,
implying for the radius $R_{1}$ the differential equation 
\begin{equation}
\frac{dR_{1} (\lambda)}{d\lambda} = -\pi R_{1} (\lambda) \ ,
\end{equation}
with the expected solution $R_{1} (\lambda) =R e^{-\pi \lambda}$.

\section{Boundary symmetries}
\setcounter{equation}{0}

Up to now we have discussed the perturbed theory from the point of
view of the bulk. In particular, we have analysed whether
gluing conditions of the chiral bulk fields may be adjusted upon a
bulk  deformation. Obviously every bulk field for which we can find a
gluing condition gives rise to a symmetry of the boundary
theory. However, as we shall see, the converse is not strictly true. 

In the following we shall therefore study the boundary symmetries
directly from the point of view of the boundary theory. The symmetry
algebra of the  
boundary theory is described by the set of `holomorphic' fields $S$
(with integer conformal weight) that are local with respect to all
other boundary fields (in the sense that there are no branch cuts in
the boundary OPEs). The full spectrum of boundary fields then forms a
representation of this algebra. In a weaker sense, we can think of a
boundary symmetry as an algebra of boundary fields that are only
mutually local (but not necessarily local with respect to all the
other boundary fields). This condition guarantees that these fields
define a (conventional) vertex operator algebra. It is then clear that
the full boundary spectrum also forms a representation of this
vertex operator algebra, but in general the representations that
appear in the spectrum will be twisted rather than the usual untwisted 
representations. (Twisted representations are characterised by the
property that the monodromy of the operator in the vertex operator
algebra, when taken around the field in question, may not be trivial;
the general theory of twisted representations of vertex operator
algebras has been developed in \cite{DLM}.) 

To study the behaviour of the boundary symmetries under bulk
deformations we thus need to determine changes in the boundary
operator product expansion of two boundary fields. From this we will
be able to read off changes in the conformal weights as well as in the
locality properties of these boundary operators. For simplicity we
shall only analyse current-current deformations $\Phi =J\bar{J}$ for
which $\bar{J}=\omega(J)$ at the boundary. The results we shall obtain
(see the following section) can then be summarised as
follows:\footnote{To prove 
that the conditions in (ii) and (iii) are necessary we have assumed
that the perturbation is hermitian.}
\begin{list}{(\roman{enumi})}{\usecounter{enumi}}
\item A boundary field $S$ retains its conformal weight if 
and only if $J_{0}\, \omega(J)_{0}S=0$.
\item A boundary field $S$ retains its conformal weight and remains
self-local if and only if $J_{0}S=0$ or $\omega(J)_{0}S=0$. 
\item Two mutually local fields $S_{1},S_{2}$ that satisfy (i) and
(ii) remain mutually local to one another if and only if either  
$J_{0}S_{1}=J_{0} S_{2}=0$ or 
$\omega(J)_{0}S_{1}=\omega(J)_{0} S_{2}=0$.
\item A field $S$ continues to be local with respect to all
other boundary fields if $J_{0}S=0$ and $\omega (J)_{0}S=0$. 
\end{list}
Note that $\omega(J)_0 S=0$ is equivalent to $J_0\, \omega^{-1}(S)=0$
since $\omega$ is an automorphism of the chiral algebra. 

Case (iv) describes the strongest situation in which $S$ remains a
true symmetry of the boundary theory. If $S$ arises from the gluing of
bulk fields, then condition (iv) coincides with what we obtained in
the last section, and the algebra of local fields is precisely the
algebra $\mathcal{A}_{\text{c}}=\mathcal{A}_{\text{inv}} \cap \omega
(\mathcal{A}_{\text{inv}})$. [Note that $\mathcal{A}_{\text{c}}$
consists of those fields $S$ for which $J_0 S=0$ and 
$J_0 \, \omega^{-1}(S)=0$.] On the other hand, we see that the
condition to have self-local fields (condition~(ii)) or an algebra
of mutually local fields (condition~(iii)) is weaker. In particular
the boundary has a symmetry algebra (in the above weak sense) that may
be larger than what we get from gluing bulk symmetries. For example,
if in the unperturbed theory the boundary preserves the full chiral
algebra $\mathcal{A}$ of the bulk, then the set of boundary fields 
$\{S\in\mathcal{A},J_{0}S=0\}$ forms an algebra of mutually local
fields that is isomorphic to the algebra $\mathcal{A}_{\text{inv}}$
which is preserved in the bulk. In this sense, the boundary preserves
the same symmetries as the bulk under the deformation. We shall
illustrate the different conditions and their interpretation in an
explicit example in section~\ref{ss:su2xsu2}. We shall also see 
that these results have interesting implications for the structure of
the open string moduli space; this will be explained in
section~\ref{open}.

\subsection{The deformed boundary OPE}

Let us now study the deformed OPE of two boundary fields $S_{1} (x_{1})$ and
$S_{2} (x_{2})$. To this end we insert these two fields into arbitrary
perturbed correlators and look for singularities in $x_{1}-x_{2}$. To first
order, the change in the OPE arises from the term 
\begin{equation}\label{deformedboundaryOPE}
\int_{\mathbb{H}^{+}} d^{2}w \, S_{1}(x_{1}) S_{2} (x_{2}) \Phi
(w,\bar{w}) \ ,
\end{equation}
where the integral is regulated by the prescription 
$\Im w > \epsilon/2$. A change of the relative locality is indicated
by a logarithmic term $\log (x_{1}-x_{2})$.   

As before we assume that $\Phi = J \bar{J}$ with $\bar{J} = \omega(J)$
at the boundary. Furthermore, the OPE of $J$ and $\omega(J)$ does not
have a simple pole (since otherwise the perturbation will induce a
non-trivial RG flow at the boundary); it is thus of the form
\begin{equation}
J(w)\, \omega(J)(\bar w) \sim \frac{C}{(w-\bar w)^2} + {\cal O}(1)  
\ , 
\end{equation}
with some constant $C$.

By the usual recursive procedure, we can evaluate (\ref{deformedboundaryOPE}) 
by using the singular part of the OPE of $J(w)$ with the
other fields. Since the singular term with $\bar{J}$ is independent of
$x_1$ and $x_2$, it does not give rise to the term of interest. The
other two OPEs on the other hand lead to
\begin{multline}
\int_{\mathbb{H}^+}d^2w\, \Bigg[ \sum_{m=0}^{h_{S_{1}}} 
\frac{1}{(w-x_1)^{m+1}} \,
 V(J_m S_{1},x_1) \, S_{2}(x_2)
\, \omega(J)(\bar w)  \\
+ \sum_{m=0}^{h_{S_{2}}} \frac{1}{(w-x_2)^{m+1}}\, 
S_{1}(x_1)\, V(J_m S_{2},x_2)
\,  \omega(J)(\bar w) \Bigg]
\ . \label{bubd1}
\end{multline}
We then apply the same recursive procedure for $\omega(J)$. For each of
the above two terms there are in turn two terms, where the OPE of 
$\omega(J)$ with the fields at $x_1$ and $x_2$ is considered.
Since we are only interested in a contribution proportional to
$\log(x_1-x_2)$, only the mixed terms can contribute 
\begin{multline}
\int_{\mathbb{H}^+}d^2w\, \sum_{m=0}^{h_{S_{1}}}\sum_{n=0}^{h_{S_{2}}}
\Bigg( V(J_m S_{1},x_1)\; V(\omega(J)_n S_{2},x_2)
\frac{1}{(w-x_1)^{m+1}} \frac{1}{(\bar w - x_2)^{n+1}} \\
+ V(\omega(J)_m S_{1},x_1)\; V(J_n S_{2},x_2) 
\frac{1}{(\bar w-x_1)^{m+1}} 
\frac{1}{(w-x_2)^{n+1}} \Bigg)
\ . \label{bubd2}
\end{multline}
To evaluate the integrals, let us first consider the terms with
$n>0$. By the familiar trick the integral can then be rewritten as an 
integral over the real axis,  
\begin{align}
\int_{\mathbb{H}^+} d^2w\, \frac{1}{(w-x_1)^{m+1}} 
\frac{1}{(\bar w-x_2)^{n+1}} & = 
\frac{i}{2n}\int_{\mathbb{R}+i\epsilon/2}\! dw \, \frac{1}{(w-x_1)^{m+1}} 
\frac{1}{(\bar w-x_2)^n}\nonumber\\
& = \frac{i}{2n}\int_{\mathbb{R}} dx \,
\frac{1}{(x-x_1+i\frac{\epsilon}{2})^{m+1}}  
\frac{1}{(x-x_2-i\frac{\epsilon}{2})^n}\ .
\end{align}
As $n>0$, the integral falls off fast enough so that we can close the
contour in the lower half plane. By the residue theorem, the result is
then some inverse power of $x_1-x_2$, but certainly not
logarithmic. The same argument applies for $m>0$, thus we can
concentrate on $m=n=0$.
 
\noindent In this case we have
\begin{align}
\int_{\mathbb{H}^+} dxdy \, \frac{1}{x+iy-x_1}\frac{1}{x-iy-x_2} & = 
\int_{\epsilon/2}^\Lambda dy \, \frac{2\pi i}{2iy+(x_2-x_1)} \\
& = \pi \log (2iy + (x_{2}-x_{1})) \Big|_{\epsilon/2}^{\Lambda} \ ,
\end{align}
where we have introduced an infrared cut-off $\Lambda$.  For
$\Lambda\rightarrow\infty$, the corresponding term is independent of
$x_1, x_2$ and thus harmless. The $\epsilon$-term however produces a
(real) logarithmic term in $x_1-x_2$. As the second integral in
(\ref{bubd2}) is just the complex conjugate of the first, the
condition that $S_{1}$ and $S_{2}$ stay mutually local is therefore
\begin{equation}\label{rellocality}
V(J_0 S_{1},x_1)\; V(\omega(J)_0 S_{2},x_2) 
+ V(\omega(J)_0 S_{1},x_1)\; V(J_0 S_{2},x_2) = 0\ .
\end{equation}
This should hold inside arbitrary correlators.

We can now use this result to derive the conditions (i) -- (iv) from
the beginning of this section. A boundary field $S=S_1$ will change
its conformal weight if there exists a boundary field $S_2$ (its
conjugate field) for which the two-point function picks up a
logarithmic term. For this to be absent we therefore require that the
vacuum  expectation value of~\eqref{rellocality} vanishes. By a usual
contour deformation argument we may move the zero mode acting on the
field at $x_2$ to the field at $x_1$; since $J_0$ and $\omega(J)_0$
commute (because $J$ and $\omega(J)$ do not have a simple pole), we then
obtain the condition   
\begin{equation}
2 \langle V(J_0 \, \omega(J)_0 S,x_1) \, V(S_2,x_2) \rangle = 0 \ . 
\end{equation}
Since the two-point function on the boundary is non-degenerate, this
can only be the case for all $S_2$ if $J_0 \, \omega(J)_0 S=0$, thus
proving (i).

A boundary field $S$ will stay in addition self-local, if the OPE of
$S$ with itself does not have any logarithmic coefficients. This must
be the case in arbitrary correlators. Thus the condition means that 
\eqref{rellocality} with $S=S_1=S_2$ must vanish as an operator
identity. By considering the OPE with an arbitrary field $V(\phi,z)$
in the limit where $z\rightarrow x_1$ and $z\rightarrow x_2$, this can
only be the case if either  $J_{0}S=0$ or $\omega (J)_{0}S=0$, or if 
$J_{0}S$ and $\omega(J)_0 S$ are proportional to one another (so that
the two terms in~\eqref{rellocality} cancel). In the last case, using
(i), it follows that $J_0 J_0 S = 0$ and 
$\omega(J)_0\, \omega(J)_0S=0$. If the perturbation is hermitian, 
{\it i.e.} if both $J_0$ and $\bar{J}_0$ are self-adjoint operators, 
we may diagonalise $J_0$. Then $J_0 J_0 S = 0$ implies that 
$J_0 S=0$, and hence either $J_{0}S=0$ or $\omega(J)_{0}S=0$, thus
proving (ii). 

Now consider two mutually local fields $S_{1}$ and $S_{2}$ that
satisfy (i) and (ii). The condition that they remain mutually local
is again that \eqref{rellocality} holds inside an
arbitrary correlator, {\it i.e.}\ as an operator identity. Because of
(ii), either $J_{0}S_1=0$ or $\omega (J)_{0}S_1 =0$, and either 
 $J_{0}S_2=0$ or $\omega (J)_{0}S_2 =0$. It is then obvious that
\eqref{rellocality} only vanishes if either 
$J_{0}S_1=J_0 S_2 =0$ or $\omega (J)_{0}S_1 = \omega (J)_{0}S_2 =0$, 
thus giving (iii).

Finally if we want a field $S$ to stay local relative to all fields
$S'$, \eqref{rellocality} must hold as an operator identity for
$S_1=S$ and $S_2=S'$ arbitrary. This is obviously the case if 
$J_0 S = \omega (J)_{0}S=0$. Thus we obtain (iv).
\medskip

All the given arguments can be generalised to higher orders; in
appendix~B this is explained for the analysis of (ii), but the line of
arguments is similar in the other cases.

\subsection{Open string moduli space}\label{open}

These considerations also have some interesting 
implications for the structure of the open
string moduli space. The moduli space is spanned by the exactly
marginal boundary fields. The fields $S$ that keep conformal weight
$h=1$ and thus stay marginal under the bulk deformation satisfy
$J_{0}\omega (J)_{0}S=0$. This does not guarantee however that the
fields remain {\em exactly} marginal. As was shown in
\cite{Recknagel:1998ih} a sufficient criterion for exact marginality
is that the marginal field is self-local. The criterion for
self-locality (ii) thus provides a characterisation of at least some
of the exactly marginal boundary fields.

On the other hand, at least to first order in perturbation theory, the
condition for exact marginality of $S$ is only that no non-trivial
relevant or marginal fields appear in the OPE of $S$ with
itself. Thus the space of exactly marginal boundary fields (that
parameterise the open string moduli space) could be bigger than just
the self-local marginal fields. Actually, from the above analysis of
the perturbed OPE  it is clear that the only
modification of the OPE of a marginal field $S$ with itself appears in
the form of terms containing a logarithm, which implies that the only
effect is to change the conformal weights of the fields that appear in
the OPE. If in the original theory the only fields that appear in the
OPE of $S$ with itself have $h>1$, then this will continue to be so,
at least for some finite range of $\lambda$. In particular, one should
therefore expect that exactly marginal boundary fields that retain
their conformal weight $h=1$ under the deformation ({\it i.e.}\ that
satisfy (i)) will continue to be exactly marginal for finite, but
maybe small $\lambda$. In general there may thus be more exactly
marginal fields than those characterised by (ii); we shall see an
example of this phenomenon in section~4.3.  \smallskip

One can also arrive at this conclusion from a different point of
view. Suppose that $S$ is an exactly marginal boundary field before
the deformation. Since we are only considering bulk deformations
$\Phi$ that are exactly marginal (in the presence of the boundary), we
know that no marginal or relevant boundary fields appear in the
bulk-boundary OPE of $\Phi$. If the direction in the open string
moduli space corresponding to $S$ survives the bulk deformation, then
$\Phi$ must also remain exactly marginal with respect to the perturbed
boundary condition. To analyse the bulk-boundary OPE
of $\Phi$ in the deformed boundary theory, we look at the deformed
correlators containing $\Phi(w,\bar{w})$ and look for singularities in
$w-\bar{w}$. The first order contribution is
\begin{equation}
\Phi (w,\bar{w}) \int_{\mathbb{R}} dx \, S (x) \ ,
\end{equation}
where, as usual, the expression is understood as being inserted into
correlators. Writing $\Phi =J\bar{J}$, and using the OPE of $J$ and
$\omega (J)$ with $S$, we see that the only terms that could change
the singular terms in $w-\bar{w}$ are
\begin{equation}
\int_{\mathbb{R}} dx\, \sum_{\substack{m,n\geq 0\\
m+n\leq 1}} V (\omega (J)_{n}J_{m}S,x) \frac{1}{(w-x)^{m+1}
(\bar{w}-x)^{n+1}} \ .
\end{equation}
Each summand gives a contribution $\sim (w-\bar{w})^{-m-n-1}$, so the
only problematic term is the one with $m=n=0$. For this term to be
absent, we need that $\omega (J)_{0} J_{0}S=0$.  This coincides with
the condition (i) that $S$ does not change its conformal weight under
the bulk deformation $\Phi$.

\subsection{Example: Deformed $SU (2)\times SU (2)$ permutation branes}
\label{ss:su2xsu2}

Let us illustrate our analysis in an example. We consider the product
of two $SU (2)$ WZW models at level $k$. On the upper half plane we
impose permutation gluing conditions on the currents
(see~\cite{Figueroa-O'Farrill:2000ei,Recknagel:2002qq}) corresponding
to the automorphism 
\begin{equation}
\omega(J^{(1)}) = g\, J^{(2)} \, g^{-1} \ , \qquad 
\omega(J^{(2)}) = h\, J^{(1)} \, h^{-1} \ ,  
\end{equation}
where $g$ and $h$ are group elements in $SU(2)$. Since this
gluing condition preserves the full $su(2)_k \times su(2)_k$ symmetry,
the boundary spectrum of each permutation brane forms a (conventional)
representation of $su(2)_k\times su(2)_k$. In the simplest case (the
brane associated to the identity representation with $g$ and $h$ being
arbitrary) the spectrum takes the form  
\begin{equation}\label{unpbou}
\mathcal{H} = \bigoplus_{l=0}^{k} \mathcal{H}^{su (2)_{k}}_{l} \otimes
\mathcal{H}^{su (2)_{k}}_{l}\ .
\end{equation}
Now we perturb the theory by the operator 
$\Phi =J^{(1)}_{3}\bar{J}^{(1)}_{3}$, \ie we deform\footnote{The
deformation of untwisted D-branes in a single copy of $SU(2)$ has been
analysed in~\cite{Forste:2001gn,Forste:2003ne,Fredenhagen:2006dn}.}
the $U(1)$ sitting inside the first $SU (2)$. The symmetry in the bulk
is broken to 
\begin{equation}\label{bulksy}
\mathcal{A}_{\text{inv}}=su(2)_{k}/u(1)_{2k} \times u (1) \times
su(2)_{k}\ . 
\end{equation}
The chiral bulk symmetry thus only contains four fields of conformal
weight one after the perturbation; these are $J_3^{(1)}$ and 
$J_a^{(2)}$. We also note that the bulk deformation is exactly
marginal in the presence of the permutation boundary, because the OPE
of $J^{(1)}_{3}$ and  $\omega(J^{(1)}_{3})=g J^{(2)}_{3} g^{-1}$ is
non-singular.  

As we have just seen, the chiral and anti-chiral fields 
$J^{(1)}_{\pm}$ and $\bar{J}^{(1)}_{\pm}$ do not remain chiral under
the deformation. After the deformation we therefore cannot glue 
$J^{(1)}_{\pm}$ to $h^{-1} \bar{J}^{(2)}_{\pm} h$ and 
$\bar{J}^{(1)}_{\pm}$ to $gJ^{(2)}_{\pm}g^{-1}$ any more. The chiral
algebra that can still be preserved by gluing bulk fields is therefore
only  
\begin{equation}\label{Ap}
\mathcal{A}_{\text{c}}=\mathcal{A}_{\text{inv}}\cap 
\omega (\mathcal{A}_{\text{inv}}) = 
\frac{su (2)_{k}}{u (1)_{2k}}\times u(1) \times 
\frac{su (2)_{k}}{u(1)_{2k}} \times u (1) \ . 
\end{equation}
Thus we are in the situation where 
$\mathcal{A}_{\text{c}} \subsetneq \mathcal{A}_{\text{inv}}$. Note that 
$\mathcal{A}_{\text{c}}$ has only two fields of conformal weight one. 
\medskip

Now we turn to the boundary description of the system. The boundary 
fields that belong to the $SU (2)$ currents $J^{(1)}_{\pm}$ and
$gJ^{(2)}_{\pm}g^{-1}$ keep their conformal weight: they satisfy
criterion (i) of section~4, because they are either annihilated by 
the zero mode of 
$J^{(1)}_{3}$ or by the zero mode of
$\omega(J^{(1)}_{3})=gJ^{(2)}_{3}g^{-1}$. The boundary theory
therefore continues to have six marginal fields. Furthermore, since
they are exactly marginal in the original theory, they remain exactly
marginal, at least for some finite perturbation. (As we shall see
momentarily, they will actually remain exactly marginal for arbitrary
finite perturbations.) This is in agreement with the 
arguments of section~4.2; the six-dimensional moduli space of
permutation branes should survive the perturbation, because the bulk
deformation $\Phi$ is exactly marginal for arbitrary $g$ and $h$.
In the perturbed theory, these six degrees of freedom can be described
as follows: two parameterise the choice of gluing $J^{(1)}_{3}$ to any
current of the second, undeformed $su(2)$, similarly two come from
gluing $\bar{J}^{(1)}_{3}$, and the remaining two come from the two $u
(1)$s that are conserved by the brane.

Not all of these fields are however mutually local, and therefore
arbitrary linear combinations will not be self-local. Given the
analysis of section~4 we expect that a subalgebra of fields isomorphic
to $\mathcal{A}_{\text{inv}}$ remains mutually local. For example we
can take the mutually local fields to be those that are annihilated by
the zero mode of $J^{(1)}_3$. This subalgebra contains then four
fields of $h=1$, namely $J^{(1)}_3$ as well as the three fields
$J^{(2)}_a$  from the second $su(2)_k$.

Finally, the fields that are local with respect to all boundary fields
are those that are annihilated by both the zero mode of $J^{(1)}_3$
and $J^{(2)}_3$; this algebra is then precisely
$\mathcal{A}_{\text{c}}$ and contains 
only two fields of weight one, namely $J^{(1)}_3$ and $J^{(2)}_3$.
\medskip

We can check these assertions by computing the boundary spectrum. The
deformed theory is a $\mathbb{Z}_{k}\times \mathbb{Z}_{k}$ orbifold of
the product of two parafermion theories $su (2)_{k}/u (1)_{2k}$ and a
square torus, with permutation gluing conditions in the coset part and
a diagonal one-dimensional brane on the torus (similar to the
situation in section~3.3). The permutation boundary state on the
parafermions is not affected by the perturbation, and it is
straightforward to determine the boundary states for the diagonal
branes on the deformed torus. It is then not hard to obtain the
boundary states in the orbifolded theory and from there the boundary
spectra.

We shall take a shorter route here which uses geometric arguments.
First we decompose the boundary spectrum (\ref{unpbou}) of the
unperturbed theory into representations of $su (2)_{k}/u (1)_{2k}$ and
$u (1)_{2k}$,
\begin{equation}
\mathcal{H} = \bigoplus_{l,m_{1},m_{2}} \mathcal{H}_{(l,m_{1})}\otimes
\mathcal{H}_{(l,m_{2})}\otimes \mathcal{H}^{u (1)_{2k}}_{m_{1}} \otimes
\mathcal{H}^{u (1)_{2k}}_{m_{2}}\ ,
\end{equation}
where $m_{1},m_{2}$ run from $-k+1$ up to $k$ (with the condition
$l+m_{i}$ even). 
The product of the $u (1)_{2k}$ representations can be explicitly
expressed in terms of momentum and winding modes. On the level of
characters we have ($m_{1}+m_{2}$ is even)
\begin{equation}
\chi_{m_{1}} (q) \chi_{m_{2}} (q) = \frac{1}{\eta^{2} (q)} 
\sum_{n\in\mathbb{{Z}}} 
\sum_{m=\frac{m_{1}+m_{2}}{2}+nk\ \text{mod}\, 2k}
q^{\frac{m^{2}}{2k} + (n-\frac{m_{1}-m_{2}}{2k})^{2}\frac{k}{2}} \ .
\end{equation}
Here, $m$ corresponds to the momentum modes of the open strings, and
$n$ corresponds to the winding modes.  When we deform the radius of
the first $U (1)$, $R_{1}\to \kappa R_{1}$, the contribution
of momentum and winding modes change. The length of the brane is
changed to $\sqrt{R_{1}^{2}+R_{2}^{2}}$, which means that the conformal
weight of a momentum mode is changed by the factor $2/
(1+\kappa^{2})$. A string winding perpendicular to the brane has
length $R_{1}R_{2}/\sqrt{R_{1}^{2}+R_{2}^{2}}$, so the conformal
weight of a winding mode is changed by a factor of
$2\kappa^{2}/(1+\kappa^{2})$ (see figure~\ref{fig:branelength}). Hence
the boundary partition function of the permutation brane in the
perturbed theory is
\begin{equation}\label{partfunc1}
Z (q) = \sum_{l,m_{1},m_{2}} \sum_{n}
\sum_{m=\frac{m_{1}+m_{2}}{2}+nk\ \text{mod}\, 2k}
\chi_{(l,m_{1})} (q) \chi_{(l,m_{2})} (q) \frac{1}{\eta^{2} (q)}
q^{\frac{m^{2}}{k (1+\kappa^{2})}
+(n-\frac{m_{1}-m_{2}}{2k})^{2}\frac{k\kappa^{2}}{(1+\kappa^{2})}} \ . 
\end{equation}
\begin{figure}
\begin{center}
\input{branelength.pstex_t}
\end{center}
\caption{\label{fig:branelength}The diagonal brane with length $L_{b}$
and a string with length $L_{s}$ that winds perpendicular to the brane
around the torus.}
\end{figure}
In particular, the partition function can be written in terms of
untwisted representations of $\mathcal{A}_{\text{c}}$, as
anticipated. Given our analysis we expect, however, that we can also
write the partition function in terms of (twisted) representations of
$\mathcal{A}_{\text{inv}}$. To see that this is indeed possible we
rewrite the partition function as
\begin{equation}
Z (q) = \sum_{l,m_{1},m_{2}} \sum_{N,M} \chi_{(l,m_{1})} (q)
\chi_{(l,m_{2})} (q) \frac{1}{\eta^{2} (q)}
q^{\big[(\frac{m_{1}+m_{2}}{2k}+M+N)^{2} +
(N-\frac{m_{1}-m_{2}}{2k}-M)^{2} \kappa^{2}
\big]\frac{k}{1+\kappa^{2}}}\ ,
\end{equation}
where we introduced new summation variables $M,N$ which replace the
variables $m,n$ in~\eqref{partfunc1} by
$m=\frac{m_{1}+m_{2}}{2}+ (N+M)k$, $n=N-M$. 
A simple transformation of the exponent of $q$ yields now
\begin{equation}
Z (q) = \sum_{l,m_{1},m_{2}} \sum_{N,M} \chi_{(l,m_{1})} (q)
\chi_{(l,m_{2})} (q) \frac{1}{\eta^{2} (q)}
q^{\frac{(m_{2}+2Nk)^{2}}{4k} + \frac{(m_{1}+2Mk)^{2}}{4k} +
\frac{1}{2k}(m_{2}+2Nk) (m_{1}+2Mk)\frac{1-\kappa^{2}}{1+ \kappa^{2}}}
\ .
\end{equation}
The only effect of the deformation $\kappa\not= 1$ is the term that
is a product of the two $u(1)$ charges. Introducing twisted $su(2)$
characters,
\begin{equation}
\hat\chi_{l,\theta} (q) = \sum_{m\in\mathbb{{Z}}} \chi_{(l,m)} (q)
\frac{1}{\eta (q)} q^{\frac{m^{2}}{4k}+\theta m} \ ,
\end{equation}
we can rewrite the partition function as
\begin{equation}
Z (q) = \sum_{l,m_{1}} \sum_{M} \chi_{(l,m_{1})} (q)\,
\hat\chi_{l,(M+\frac{m_{1}}{2k})\frac{1-\kappa^{2}}{1+\kappa^{2}}} (q)
\, \frac{1}{\eta (q)} \, q^{\frac{(m_{1}+2Mk)^{2}}{4k}} \ .
\end{equation}
Thus the partition function can indeed be written in terms of 
twisted $su(2)$ characters, coset characters and $u(1)$ characters,
giving strong support to our claim that the boundary spectrum forms a
(twisted) representation of $\mathcal{A}_{\text{inv}}$. 

Finally, we want to check our assertion that the two sets of $SU(2)$
currents remain marginal (and in fact exactly marginal). For
concreteness let us consider the $SU (2)$ current $J^{(1)}_{+}$ that 
appears in the representation with $l=0$, $m_{1}=2$, $m_{2}=0$, $n=0$,
$m=1$. We can easily evaluate the conformal weight of this mode in the
perturbed theory, 
\begin{equation}
h = h_{(0,2)}+h_{(0,0)}+\frac{1}{k (1+\kappa^{2})} +
\frac{\kappa^{2}}{k (1+\kappa^{2})} = 1 \ ,
\end{equation}
where $h_{(0,2)}=1-\frac{1}{k}$ and $h_{(0,0)}=0$. The analysis is
similar for $J^{(1)}_{-}$ and $J^{(2)}_{\pm}$. Thus it follows that
for all values of $\kappa$, there are six marginal fields which come
from the $SU (2)\times SU (2)$ symmetry of the undeformed theory. As
we have explained before they describe the six exactly marginal
boundary fields that generate the six-dimensional moduli space of the
brane.

\subsection{Matrix factorisation examples}

As a final example we consider theories for which the perturbation is
not a current-current deformation.  While we cannot study these 
cases in general, there is an interesting class of examples for which
we can test the above ideas. These are B-type branes in Gepner models
of Calabi-Yau $3$-folds discussed in section~2.2. As was explained
there, under complex structure deformations, the diagonal $N=2$
algebra continues to be a symmetry of the bulk conformal field
theory. Suppose we consider a B-type D-brane in this background. Then
we can ask whether it will continue to be a B-type $N=2$ brane under
these complex structure deformations. This problem is very accessible
since B-type branes in these theories can be described in terms of
matrix factorisations
(see~\cite{Kapustin:2002bi,Brunner:2003dc,Kapustin:2003ga} for some
early references and~\cite{Hori:2004zd} for a review) of the
associated Landau-Ginzburg model.

Our arguments above now suggest that the brane will remain B-type
provided that the bulk perturbation is exactly marginal in the
presence of the brane, {\it i.e.}\ that no relevant or marginal
boundary field is induced by the bulk perturbation. As in the
situation discussed in \cite{Baumgartl:2007an}, a non-trivial RG flow
will be induced if and only if the bulk-boundary operator product
coefficient $B_{\Phi \psi}\neq 0$ for some boundary fermion $\psi$ of
$u(1)$ charge $1$. Thus we want to show that the matrix factorisation 
is obstructed against the bulk perturbation by $\Phi$ if and only if
$B_{\Phi \psi}\neq 0$ for some boundary fermion $\psi$ of $u(1)$
charge $1$.

{}From a matrix factorisation point of view a D-brane remains B-type
if we can adjust the matrix factorisation $Q$ as 
$Q(\lambda)=Q + \lambda Q_1 + \cdots$ so that 
$Q(\lambda)$ is a matrix factorisation of  
$W(\lambda) = W+\lambda \Phi$. To first order in $\lambda$ 
a necessary and sufficient condition for this is that $\Phi$ is exact
with respect to $Q$ \cite{Hori:2004ja}. If $\Phi$ is exact, then the
bulk-boundary correlator $B_{\Phi \psi}$ vanishes for all  
boundary fields $\psi$, as follows by standard Landau-Ginzburg
arguments. Thus if the matrix factorisation can be adjusted as above,
the bulk-boundary correlator $B_{\Phi \psi}$ vanishes, and 
no marginal or relevant boundary field is turned on.

To show the converse direction we need to prove that if $\Phi$ is not
exact with respect to $Q_0$ (so that the matrix factorisation
adjustment is obstructed), then there is a boundary field $\psi$ of
$u(1)$ charge $1$ such that $B_{\Phi \psi}\neq 0$.  If $\Phi$ is not
exact, this means that $\Phi$ taken to the boundary induces a chiral
primary field on the boundary. From conformal field theory we then
know that there is a non-trivial bulk boundary correlator involving
these two fields; since both are chiral primary fields, this amplitude
is then also non-trivial in the topologically twisted theory, and
hence $B_{\Phi \psi}\neq 0$ for some boundary field
$\psi$. Furthermore, it follows from charge conservation considerations
in the Landau-Ginzburg theory that the bulk-boundary OPE can only be
non-trivial if $\psi$ is a fermion of charge $1$. This then proves our
claim.  
\medskip

As an aside we note that the above analysis only applies to the
Calabi-Yau $3$-fold case. For the case of K3 the charge conservation
analysis implies that the only boundary field that can have
a non-trivial bulk-boundary correlator with the complex structure
deformation $\Phi$ is a boson of charge $0$. Such a field does 
{\it not} correspond to an exactly marginal boundary field. However,
this does not invalidate our claim concerning the symmetries of the
boundary since in the K3 case there is in fact $N=4$ supersymmetry,
and a brane need not remain B-type with respect to any $N=2$
subalgebra even if it continues to preserve the full $N=4$
superconformal algebra --- see \cite{Brunner:2006tc} for an
example. Finally, we also note that the above point of view 
suggests how the criterion of \cite{Brunner:2006tc} for obstructions
of matrix factorisations on K3 against complex structure deformations
can be sharpened: the matrix factorisation will be obstructed to first
order if and only if $B_{\Phi \psi}\neq 0$ for some boundary boson of
charge $0$. One easily checks that the examples of
\cite{Brunner:2006tc} are in agreement with this criterion.

\section{Conclusions}

In this paper we have analysed bulk and boundary symmetries in
perturbed conformal field theories where the perturbing field is
exactly marginal. As one may expect, generically the chiral symmetry 
algebra $\mathcal{A}$  of the bulk theory will be reduced, and we have
given a precise condition for which generators survive the
deformation, see (\ref{chiralcond}). For the case of a current-current
deformation, $\Phi=J\bar{J}$, the condition is very simple: the
surviving symmetry algebra $\mathcal{A}_{\text{inv}}$ is formed by the
fields $S\in \mathcal{A}$ for which $J_{0}S=0$. 

We have also analysed whether the symmetries of a boundary condition
are preserved under the deformation by a current-current
deformation. We have seen that if the boundary condition preserves  
$\mathcal{A}_{\text{inv}}$ before deforming the bulk, it always does so
after the deformation. Thus the boundary never destroys any additional
symmetries!

As we have seen the statement about the boundary symmetries is
actually quite subtle, and two cases need to be distinguished.
In the first case the boundary condition originally preserves the 
algebra $\mathcal{A}_{\text{inv}}$ in the sense 
that the fields in $\mathcal{A}_{\text{inv}}$ and
$\bar{\mathcal{A}}_{\text{inv}}$ are glued at the boundary via  
some automorphism $\omega$ (that is an automorphism of 
$\mathcal{A}_{\text{inv}}$). Then the symmetries of the boundary
theory actually arise from gluing preserved bulk symmetries (see
section~3.2). The prime example for this phenomenon is the 
diagonal brane on a torus when the torus is deformed (see
section~3.3). 

In the second case, the automorphism $\omega$ of $\mathcal{A}$ actually
does not define an automorphism of the preserved subalgebra 
$\mathcal{A}_{\text{inv}}$. Then only fields in
$\mathcal{A}_{\text{c}}=\mathcal{A}_{\text{inv}}\cap
\omega(\mathcal{A}_{\text{inv}})$ can be glued after the
deformation. From the point of view of the gluing conditions of the
bulk the boundary symmetry thus appears to be reduced to
$\mathcal{A}_{\text{c}}$. However, as discussed in section~4, the
boundary theory still has mutually local boundary fields associated to
all elements of $\mathcal{A}_{\text{inv}}$, and the spectrum of the
boundary theory can be decomposed into (twisted) representations of
$\mathcal{A}_{\text{inv}}$. The prime example for this phenomenon is
the permutation brane in the product of two SU(2) WZW models (see
section~4.3). 

These results were obtained for current-current deformations, but we
believe that the general observation --- the boundary symmetry is not
further reduced than the bulk symmetry --- is also true for a larger
class of deformations. We gave one example in section~4.4 using
techniques from matrix factorisations in Landau-Ginzburg models.

The appearance of boundary symmetry algebras that are not local with
respect to all boundary fields (and that therefore cannot arise from
gluing chiral bulk fields) might have an interesting consequence for
the construction of new boundary theories. The standard strategy for
how to obtain boundary states in rational conformal field theories is
by requiring gluing conditions for the chiral algebra such that this 
algebra is also present on the boundary. However, our analysis shows
that the boundary theory can have a bigger symmetry than that which is
inherited from the bulk. A similar phenomenon was already noted in a
specific example in \cite{Fredenhagen:2006qw} (see the spectrum of the
$P$-branes in equation~(2.37) in that reference) in trying to
construct the generalised permutation branes of
\cite{Fredenhagen:2005an} in conformal field theory. One may also
suspect that this idea may be relevant for the additional rank 1
factorisations that involve `non-consecutive' factors and do not
correspond to permutation branes \cite{Brunner:2005fv}: the matrix
factorisation point of view suggests that their boundary symmetry 
is as large as that of the permutation branes, but it is clear that
this cannot arise from gluing symmetries in the bulk.

In this paper we have only considered current-current deformations 
of the form $\Phi=J\bar{J}$ with $\omega(J)=\bar{J}$ at the boundary.
It is straightforward to generalise our analysis to the more general
case where $\Phi =J\bar{J}'$ with $\omega(J')=\bar{J}'$ at the
boundary. We then have to distinguish 
$\mathcal{A}_{\text{inv}}=\{S;J_{0}S=0 \}$ and 
$\bar{\mathcal{A}}'_{\text{inv}}=\{\bar{S}; \bar J'_{0}\bar{S}=0\}$. The 
gluing automorphism can be deformed for fields in
$\mathcal{A}_{\text{inv}}\cap \omega (\mathcal{A}'_{\text{inv}})$, and
$\mathcal{A}_{\text{inv}}$ still plays the role of an algebra of
mutually local boundary fields. 

Many results also carry over to the most general current-current
deformation where $\Phi = \sum_i J_i \bar{J}'_{i}$. This deformation
is exactly marginal in the bulk if the currents $J_{i}$ do not have
simple poles among themselves, and similarly for the
$\bar{J}'_{i}$~\cite{Chaudhuri:1988qb}. Using the same arguments as in
section~2 one easily shows that the symmetry algebra that is preserved
by such a deformation is $\mathcal{A}_{\text{inv}}=\{S;J_{i,0}S=0 \
\text{for all}\ i \}$ and similarly
$\bar{\mathcal{A}}'_{\text{inv}}=\{\bar{S};\bar{J}'_{i,0}\bar{S}=0\
\text{for all}\ i\}$. In the presence of a boundary, the deformation
$\Phi$ is exactly marginal if $\bar{J}'_{i}=\omega(J'_{i})$ at the
boundary and if the $J_{i}$ do not have simple poles with the $\omega
(J'_{j})$. In this case one can then also generalise straightforwardly
our analysis of deformed gluing conditions and boundary algebras.

\section*{Acknowledgements}

This research has been partially supported by the Swiss National
Science Foundation and the Marie Curie network `Constituents,
Fundamental Forces and Symmetries of the Universe'
(MRTN-CT-2004-005104). We thank Marco Baumgartl, Ilka Brunner,
Christian Hillmann, Ingo Runkel and Stefan Theisen for useful
discussions. 

\appendix

\section{Automorphism for current algebras}
\renewcommand{\theequation}{A.\arabic{equation}}
\setcounter{equation}{0}

In the case of current algebras, let us explicitly check that the
modified map $\omega_{\lambda}$ (\ref{eq3}) is in fact an
automorphism. Let $S^{(1)}$, $S^{(2)}$ be two currents such that
$\omega (S^{(i)})$ commute with $J$ and $\omega(J)$, so that there is
no single pole in their OPE. The effect of $\omega_{\lambda}$ on the
modes $S^{(i)}_{m}$ is to 
first order
\begin{equation}
S^{(i)}_{m} \mapsto \omega (S^{(i)}_{m}) -\pi \lambda \big( J,\omega
(S)^{(i)}\big) \omega (J)_{m} +\pi 
\lambda \big(\omega (J),\omega (S^{(i)})\big) J_{m} \ .
\end{equation}
Here, $(J,S)$ denotes the inner product of the generators
corresponding to $J$ and $S$. Now let us check that the commutation
relations of the current algebra are preserved under the map
$\omega_{\lambda}$, so we have to check whether
\begin{equation}
[\omega_{\lambda} (S^{(1)}_{m}),\omega_{\lambda} (S^{(2)}_{n})] =
\omega_{\lambda } ([S^{(1)}_{m},S^{(2)}_{n}]) \ .
\end{equation}
Evaluating the left hand side, we find
\begin{equation}
[\omega_{\lambda} (S^{(1)}_{m}),\omega_{\lambda} (S^{(2)}_{n})] =
\omega ([S^{(1)}_{m},S^{(2)}_{n}]) \ .
\end{equation}
We used here that in the commutators of $\omega (S^{(i)})$ with
$J,\omega (J)$ only central terms appear; the total central term is
then easily seen to vanish.

The commutator on the right hand side is just the one that we would
have got with the unchanged automorphism $\omega$. Therefore
$\omega_{\lambda }$ is an automorphism if it acts in the same way as
$\omega$ on the commutator $[S^{(1)}_{m},S^{(2)}_{n}]$. By definition
\begin{align}
\omega_{\lambda } ( [S^{(1)}_{m},S^{(2)}_{n}]) & = \omega
([S^{(1)}_{m},S^{(2)}_{n}]) - \pi \lambda 
\big(\omega (J),[\omega (S^{(1)}),\omega (S^{(2)})]\big) J_{m+n} 
\nonumber\\
& \qquad +\pi \lambda \big( J,[\omega
(S^{(1)}),\omega (S^{(2)})]\big) \omega (J)_{m+n} \ . \label{testa4}
\end{align}
Here, $[S^{(1)},S^{(2)}]$ denotes the Lie algebra bracket
corresponding to the currents $S^{(i)}$. Since 
$(J,[\omega(S^{(1)}),\omega (S^{(2)})])= 
([J,\omega (S^{(1)})],\omega (S^{(2)}))=0$ and similarly for the other
term, the two additional terms in (\ref{testa4}) cancel, and we have
shown that $\omega_{\lambda}$ is an automorphism.

\section{Higher order analysis of boundary locality}
\renewcommand{\theequation}{B.\arabic{equation}}
\setcounter{equation}{0}

In this appendix we shall analyse under which conditions the
self-locality of a boundary field $S$ remains unaffected by higher
order perturbations. This continues the analysis of (ii) in
section~4.1 to higher orders.

\noindent Suppose that $J(w)\bar J(\bar w)$ is exactly marginal on the
disk, \ie that
\begin{equation}
J(w)J(z) = \frac{1}{(w-z)^2} + {\cal O}(1) \ , \qquad 
\bar{J}(\bar{w}) \bar{J}(\bar{z}) = \frac{1}{(\bar{w}-\bar{z})^2} +
{\cal O}(1) \ , 
\end{equation}
and 
\begin{equation} 
J(w)\, \omega(J)(\bar z) = \frac{C}{(w-\bar z)^2} 
+ {\cal O}(1) \ , \label{exmarg}
\end{equation}
where $C$ is some constant (that may be zero). Moreover we assume that 
$S$ does not change its conformal weight to first order in the
perturbation. The analysis in section~4.1 then tells us that 
either $J_0 S=0$ or $\omega(J)_0 S=0$. For definiteness let us assume
in the following that the first case holds, $J_{0}S=0$.

\noindent The contribution to the deformed OPE at order $n$ is
given by
\begin{equation}
\lambda^n  S(x_1)S(x_2)\prod_{i=1}^n \int d^2 w_i \,
J(w_i)\,\omega(J)(\bar w_i) \ , \label{order_n}
\end{equation}
which as usual is understood to be inserted into arbitrary
correlators. By the usual arguments this can be evaluated by summing
the singular terms that arise from the OPE of the currents with the
other fields. Note that the integral in (\ref{order_n}) is regularised
by $ |w_i-w_j|>\epsilon$ (and of course $\Im(w_i)>\epsilon/2$). The
regularisation is obviously only important if there actually is a pole
in $w_i-w_j$; otherwise it will only lead to terms of order $\epsilon$
that we can neglect.

Let us now discuss the different pole terms that arise. In the
following we will not distinguish between $J$ and $\omega(J)$, and $w$
and $\bar w$, unless stated otherwise. We will denote the singular
contribution from the operator product expansion between operators at
$u$ and $v$ by the symbol $u\rightarrow v$; with these conventions
there are the following terms to consider:
\begin{list}{(\arabic{enumi})}{\usecounter{enumi}}
\item $w_k \rightarrow \bar w_k$: The pole $(w_k-\bar w_k)^{-2}$ only gives
a (divergent) overall renormalisation.
\item $w_k \rightarrow \bar w_j$ and $\bar w_k \rightarrow w_j$:
Again the $w_k$ and $w_j$ integrals are independent of all other
variables and thus only give an overall normalisation
factor. The same applies for $w_k \rightarrow w_j$ and 
$\bar w_k \rightarrow \bar w_j$.
\item $w_k \rightarrow w_j$ but $\bar w_k \rightarrow w_l$: Evaluate
\begin{equation} 
\int d^2w_k \, (w_k-w_j)^{-2}(\bar w_k -w_l)^{-2}=\int_{\partial
\mathbb{H^+}, w_i} dw_k\,  (w_k-w_j)^{-2}(\bar w_k-w_l)^{-1}\ ,  
\end{equation} 
where the integral is taken along the real axis, as well as around
little circles surrounding the points $w_i$. The contour integral
around $w_{i}$ with $i\neq j,l$ is zero since the function 
is regular at that point. Similarly, the integral around $w_j$ and
$w_{l}$ gives zero. The integral along the real axis
can be evaluated by closing the contour. Depending on whether $w_j$
and $w_l$ are in the upper or in the lower half plane (or more
precisely, whether we consider $w$ or $\bar w$), it gives either zero
or a term $\sim (w_l-w_j)^{-2 }$. Effectively we have thus reduced the
problem to the one where $J(w_k)$ is absent, and we are considering
the pole contribution from $w_l\rightarrow w_j$.
\end{list}
This deals with all the poles between the different currents $J$ and
$\omega(J)$. It remains to analyse the poles that involve at
least one $S$. There are two more cases to consider:
\begin{list}{(\arabic{enumi})}{\usecounter{enumi}}
\setcounter{enumi}{3}
\item $w_k\rightarrow x_i$ and $\bar w_k\rightarrow x_j$: This is the
situation discussed in section~4.1. Note that we can still use
translation invariance, as the integrand is regular at all points $w_k
= w_i$. A logarithmic term in $x_{1}-x_{2}$ could only be produced if
both zero modes $J_{0}$ and $\omega (J)_{0}$ act on the fields $S$,
but this gives zero, because we assumed $J_{0}S=0$. We thus obtain
meromorphic terms in $x_1,x_2$, and new operators of the form $V(J_m
S;x_{i})$ and $V (\omega (J)_{n}S;x_{j})$, $m>0,n\geq 0$. Obviously
$[J_0,J_n]=0$, and because of (\ref{exmarg})
$[J_0,\omega(J)_n]=0$. This means that the new fields do not have any
simple pole with the current $J$, so that we can continue to use the
same arguments as before.
\item $w_k\rightarrow x_1$, but $\bar w_k\rightarrow \bar w_l$: 
By the same arguments as in (3) above we do not get anything new if
$w_{l}\to w_{j}$, so we may assume that 
$w_l\rightarrow x_i$. If $w_l\rightarrow x_1$, then the $w_k$ and
$w_l$ integrals only depend on $x_1$ and no other variables. By
translation invariance this is just a constant factor.
\end{list}
The only remaining cases are thus $w_k\to
x_1,\bar{w}_{k}\to \bar{w}_{l}$, $w_l\to x_2$, and similar situations
with $w$ and $\bar{w}$ exchanged. Let us consider first the
contribution 
\begin{equation}
\int d^{2}w_{l}\, d^{2}w_{k}\,
[J(w_{k})S(x_1)][J(w_{l})S(x_2)]\frac{1}{(\bar w_k- \bar w_l)^2}\ ,
\end{equation}
where we denote the singular part of the OPE by square brackets.
The domain of integration is restricted by $|w_{l}-w_{k}|>\epsilon$ and
$\Im w_{l},\Im w_{k}>\frac{\epsilon}{2}$. Write $w_{l}=u_{l}+iv_{l}$
and redefine $\hat{w}_{k}= w_{k}-u_{l}$. The variable $u_{l}$ is
integrated over the real axis without any restrictions. The integrand
has poles in $u_{l}$ at $u_{l}= x_{2}- i v_{l}$ and at 
$u_{l} = x_{1}-\hat{w}_{k}$ which both lie on the lower half plane. By
closing the contour in the upper half plane we thus see that this
contribution vanishes. 

The same argument applies for the case $\bar{w}_{k}\to x_{1}$,
$w_{k}\to w_{l}$, $\bar{w}_{l}\to x_{2}$, so the only remaining
contribution that we need to consider comes from
$\bar{w}_{k}\to x_{1}$, $w_{k}\to \bar{w}_{l}$, ${w}_{l}\to x_{2}$,
which is of the form 
\begin{equation}
\int d^{2}w_{l} \, d^{2}w_{k}\,  \frac{1}{(w_{k}-\bar{w}_{l})^{2}}
\sum_{m=0}^{h_{S}} \sum_{n=1}^{h_{S}} \frac{1}{(\bar{w}_{k}-x_{1})^{m+1}}
\frac{1}{(w_{l}-x_{2})^{n+1}} V (\omega (J)_{m}S,x_{1}) V (J_{n}
S,x_{2}) \ . 
\end{equation}
Note that there is no term with $n=0$. It is not hard to see that the
integral for each summand produces a contribution $\sim
(x_{1}-x_{2})^{-(m+n)}$, so that a logarithmic term could only appear for
$m=n=0$.

In summary, we have thus shown that there will be no logarithmic terms
to any order in perturbation theory if they do not arise at first
order.


\end{document}